\documentstyle[12pt]{article}
\begin{document}
\begin{flushright}
Preprint IHEP 94-74\\
July 19, 1994
\end{flushright}
\begin{center}
{\bf \large Masses of $nS$-wave Heavy Quarkonium Levels \\from QCD Sum Rules}
\vspace*{1cm}
V.V.Kiselev\\
{\it Institute for High Energy Physics,\\
Protvino, Moscow Region, 142284, Russia,\\
E-mail: kiselev@mx.ihep.su\\
Fax: +7-095-230-23-37}
\end{center}
\begin{abstract}
Using a specific scheme of the QCD sum rules, one derives an universal
formula for the mass differences for the $nS$-levels of heavy qiarkonium.
This relation does not depend on the flavours of the heavy quarks, composing
the quarkonium, and it is
in a good agreement with the experimental mass values for the
$\psi$- and $\Upsilon$-families.
\end{abstract}

\section*{Introduction}

The QCD sum rules \cite{1} are one of the powerful tools for a description of
nonperturbative characteristics of the heavy quark bound states.
However, in the consideration of masses for basic states of heavy quarkonia,
the QCD sum rule predictions have the accuracy, that is one order of magnitude
lower than the accuracy of phenomenological potential models \cite{2}.
This is caused by that, first, the QCD sum rule calculations are made in a
finite order of the QCD perturbation theory for the Wilson's coefficients
and with a restricted set of the quark-gluon condensates, so that the results
depend on an external unphysical parameter, defining a sum rule scheme
for an averaging (the number of moment for the spectral density of
current correlator or the Borel transform parameter). Second,
one has an uncertainty in a value of the threshold $s_{th}$, discriminating the
resonant region and the hadronic continuum. Moreover, the rapidly-dropping
weight functions, defining the sum rule scheme, do not allow one to extract
an information on the contribution by the higher excitations in the quarkonium,
so that this contribution is obviously neglected.

Recently in refs.\cite{3,4,5}, one has offered the QCD sum rule scheme,
which allows one to get the scaling relation for the leptonic constants $f$
of the $S$-states of different quarkonia with the mass $M$ and the
reduced quark mass $\mu$,
\begin{equation}
\frac{f^2}{M}\;\biggl(\frac{M}{4\mu}\biggr)^2 = const.\;, \label{2}
\end{equation}
and the relation for the leptonic constants $f_n$ of excited $nS$-states of the
quarkonium
\begin{equation}
\frac{f^2_{n_1}}{f^2_{n_2}} = \frac{n_2}{n_1}\;, \label{3}
\end{equation}
independently of the flavours of heavy quarks, composing the quarkonium.

In the present paper, in the framework of the QCD sum rule scheme, offered
in refs.\cite{3,4,5}, we derive the relation for the mass differences of
$nS$-wave levels in the heavy quarkonium
\begin{equation}
\frac{M_n-M_1}{M_2-M_1} = \frac{\ln {n}}{\ln {2}}\;,\;\;\;n\ge 2\;,\label{m3}
\end{equation}
independently of the heavy quark flavours.

In Section 1 we describe the used scheme of QCD sum rules and derive relation
(\ref{m3}). In Section 2 we make the phenomenological analysis of
relation (\ref{m3}), which is in a good agreement with the experimental ratios
of mass differences in the $\psi$- and $\Upsilon$-families. In the Conclusion
the obtained results are summarized.

\section{Heavy Quarkonium Sum Rules}

Let us consider the two-point correlator functions of quark currents
\begin{eqnarray}
\Pi_{\mu\nu} (q^2) & = & i \int d^4x e^{iqx} <0|T J_{\mu}(x)
J^{\dagger}_{\nu}(0)|0>\;,
\label{1} \\
\Pi_P (q^2) & = & i \int d^4x e^{iqx} <0|T J_5(x) J^{\dagger}_5(0)|0>\;,
\end{eqnarray}
where
\begin{eqnarray}
J_{\mu}(x) & = & \bar Q_1(x) \gamma_{\mu} Q_2(x)\;,\\
J_5(x) & = & \bar Q_1(x) \gamma_5 Q_2(x)\;,\\
\end{eqnarray}
$Q_i$ is the spinor field of the heavy quark with $i = c, b$.

Further, write down
\begin{equation}
\Pi_{\mu\nu} = \biggl(-g_{\mu\nu}+\frac{q_{\mu} q_{\nu}}{q^2}\biggr) \Pi_V(q^2)
+ \frac{q_{\mu} q_{\nu}}{q^2} \Pi_S(q^2)\;,
\end{equation}
where $\Pi_V$ and $\Pi_S$ are the vector and scalar correlator functions,
respectively. In what follows we will consider the vector and pseudoscalar
correlators: $\Pi_V(q^2)$ and $\Pi_P(q^2)$.

Define the leptonic constants $f_V$ and $f_P$
\begin{eqnarray}
<0|J_{\mu}(x) |V(\lambda)> & = & i \epsilon^{(\lambda)}_{\mu}\;
f_V M_V\;e^{ikx}\;,\\
<0|J_{5\mu}(x)|P> & = & i k_{\mu}\;f_P e^{ikx}\;,
\end{eqnarray}
where
\begin{equation}
J_{5\mu}(x)  =  \bar Q_1(x) \gamma_5 \gamma_{\mu} Q_2(x)\;,
\end{equation}
so that
\begin{equation}
<0|J_{5}(x)|P>  =  i\;\frac{f_P M_P^2}{m_1+m_2}\;e^{ikx}\;, \label{9}
\end{equation}
where $|V>$ and  $|P>$ are the state vectors of $1^-$ and $0^-$
quarkonia, and $\lambda$ is the vector quarkonium polarization, $k$
is 4-momentum of the meson, $k_{P,V}^2 = M_{P,V}^2$.

Considering the charmonium ($\psi$, $\psi '$ ...) and bottomonium ($\Upsilon$,
$\Upsilon '$, $\Upsilon ''$ ...), one can easily show that the relation
between the width of
leptonic decay $V \to e^+ e^-$  and $f_V$ has the form
\begin{equation}
\Gamma (V \to e^+ e^-) = \frac{4 \pi}{9}\;e_i^2 \alpha_{em}^2\;
\frac{f_V^2}{M_V}\;,
\end{equation}
where $e_i$ is the electric charge of quark $i$.

In the region of narrow nonoverlapping resonances, it follows from
eqs.(\ref{1}) - (\ref{9}) that
\begin{eqnarray}
\frac{1}{\pi} \Im m \Pi_V^{(res)} (q^2) & = &
\sum_n f_{Vn}^2 M_{Vn}^2 \delta(q^2-M_{Vn}^2)\;,
\label{11} \\
\frac{1}{\pi} \Im m \Pi_P^{(res)} (q^2) & = &
\sum_n f_{Pn}^2 M_{Pn}^4\;\frac{1}{(m_1+m_2)^2} \delta(q^2-M_{Pn}^2)\;.
\end{eqnarray}
Thus, for the observed spectral function one has
\begin{equation}
\frac{1}{\pi} \Im m \Pi_{V,P}^{(had)} (q^2)  = \frac{1}{\pi} \Im m
\Pi_{V,P}^{(res)} (q^2)+ \rho_{V,P}(q^2, \mu_{V,P}^2)\;,
\label{13}
\end{equation}
where $\rho (q^2,\;\mu^2)$ is the continuum contribution, which is
not equal to zero at $q^2 > \mu^2$.

Moreover, the operator product expansion gives
\begin{equation}
\Pi^{(QCD)} (q^2)  = \Pi^{(pert)} (q^2)+ C_G(q^2) <\frac{\alpha_S}{\pi} G^2> +
C_i(q^2)<m_i \bar Q_i Q_i>+ \dots\;,
\label{14}
\end{equation}
where the perturbative contribution $\Pi^{(pert)}(q^2)$ is labeled, and
the nonperturbative one is expressed in the form of sum
of quark-gluon condensates
with the Wilson's coefficients, which can be calculated in the QCD
perturbative theory.

In eq.(\ref{14}) we have been restricted by the contribution of vacuum
expectation values for the operators with dimension $d =4$.
For $C^{(P)}_G (q^2)$ one has, for instance, \cite{1}
\begin{equation}
C_G^{(P)} = \frac{1}{192 m_1 m_2}\;\frac{q^2}{\bar q^2}\;
\biggl(\frac{3(3v^2+1)(1-v^2)^2}
{2v^5} \ln \frac{1+v}{1-v} - \frac{9v^4+4v^2+3}{v^4}\biggr)\;, \label{15}
\end{equation}
where
\begin{equation}
\bar q^2 = q^2 - (m_1-m_2)^2\;,\;\;\;\;v^2 = 1-\frac{4m_1 m_2}{\bar q^2}\;.
\label{16}
\end{equation}
The analogous formulae for other Wilson's coefficients can be found in
Ref.\cite{1}. In what follows it will be clear that the explicit form
of coefficients has no significant meaning for the present consideration.

In the leading order of QCD perturbation theory it has been found for
the imaginary part of correlator that \cite{1}
\begin{eqnarray}
\Im m \Pi_V^{(pert)} (q^2) & = & \frac{\tilde s}{8 \pi s^2}
(3 \bar s s - \bar s^2 + 6m_1 m_2 s - 2 m_2^2 s) \theta(s-(m_1+m_2)^2),\\
\Im m \Pi_P^{(pert)} (q^2) & = & \frac{3 \tilde s}{8 \pi s^2}
(s - (m_1-m_2)^2) \theta(s-(m_1+m_2)^2)\;,
\end{eqnarray}
where $\bar s = s-m_1^2+m_2^2$, $ \tilde s^2 = \bar s^2 -4 m_2^2 s$.

The one-loop contribution into $\Im m \Pi(q^2)$ can be included into the
consideration (see, for example, Ref.\cite{1}). However, we note that the
more essential correction is that of summing a set over the powers of
$(\alpha_s/v)$, where $v$ is defined in eq.(\ref{16}) and is a relative quark
velocity, and $\alpha_S$ is the QCD interaction constant. In Ref.\cite{1}
it has been shown that account of the Coulomb-like gluonic
interaction between the quarks leads to the factor
\begin{equation}
F(v) = \frac{4 \pi}{3}\;\frac{\alpha_S}{v}\; \frac{1}{1-\exp (-\frac{4 \pi
\alpha_S}{3 v})}\;,
\end{equation}
so that the expansion of the $F(v)$ over $\alpha_S/v \ll 1$ restores,
precisely,
the one-loop $O(\frac{\alpha_S}{v})$ correction
\begin{equation}
F(v) \approx 1 - \frac{2 \pi}{3}\;\frac{\alpha_s}{v}\; \dots \label{20}
\end{equation}
In accordance with the dispersion relation one has the QCD sum rules,
which state that, in average, it is true that, at least, at $q^2 < 0$
\begin{equation}
\frac{1}{\pi}\;\int\frac{\Im m \Pi^{(had)}(s)}{s-q^2} ds = \Pi^{(QCD)}(q^2)\;,
 \label{21}
\end{equation}
where the necessary subtractions are omitted. $\Im m \Pi^{(had)}(q^2)$ and
 $\Pi^{(QCD)}(q^2)$ are defined by eqs.(\ref{11}) - (\ref{13}) and
eqs.(\ref{14}) - (\ref{20}), respectively.
eq.(\ref{21}) is the base to develop the sum rule approach in the forms
of the correlator function moments and of the Borel transform analysis
(see Ref.\cite{1}). The truncation of the set in the right hand side of
eq.(\ref{21}) leads to the mentioned unphysical dependence of the $f_{P,V}$
values on the external parameter of the sum rule scheme.

Further, let us use the conditions, simplifying the consideration due to
the heavy quarkonium.

\subsection{Nonperturbative Contribution}

We assume that, in the limit of the very heavy quark mass, the power
corrections of nonperturbative contribution are small. From eq.(\ref{15})
one can see that, for example,
\begin{equation}
C_G^{(P)}(q^2) \approx O(\frac{1}{m_1 m_2})\;,\;\; \Lambda/m_{1,2}\ll 1\;,
\end{equation}
where $v$ is fixed,  $q^2 \sim (m_1 + m_2)^2$,
when $\Im m \Pi^{(pert)}(q^2) \sim (m_1+m_2)^2$.
It is evident that, due to the purely dimensional consideration, one can
believe that the Wilson's coefficients tend to zero as
$1/m_{1,2}^2$.

Thus, the limit of very large heavy quark mass implies that one can neglect
the quark-gluon condensate contribution.

\subsection{Nonrelativistic Quark Motion}

The nonrelativistic quark motion implies that, in the resonant region, one has,
in accordance with eq.(\ref{16}),
\begin{equation}
v \to 0\;.
\end{equation}
So, one can easily find that in the leading order
\begin{equation}
\Im m \Pi_P^{(pert)}(s) \approx  \Im m \Pi_V^{(pert)}(s) \to \frac {3 v}
{8 \pi^2} s\; \biggl(\frac{4\mu}{M}\biggr)^2\;,
\end{equation}
so that with account of the Coulomb factor
\begin{equation}
F(v) \simeq \frac{4 \pi}{3}\; \frac{\alpha_S}{v}\;,
\end{equation}
one obtaines
\begin{equation}
\Im m \Pi_{P,V}^{(pert)}(s) \simeq \frac{\alpha_S}{2} s\;
\biggl(\frac{4\mu}{M}\biggr)^2\;. \label{27}
\end{equation}

\subsection{"Smooth Average Value" Scheme of the Sum Rules}

As for the hadronic part of the correlator, one can write down for the narrow
resonance contribution
\begin{eqnarray}
\Pi_V^{(res)}(q^2) & = & \int \frac{ds}{s-q^2}\;\sum_n f^2_{Vn} M^2_{Vn}
\delta(s-M_{Vn}^2)\;,
\label{28} \\
\Pi_P^{(res)}(q^2) & = & \int \frac{ds}{s-q^2}\;\sum_n f^2_{Pn}
\frac{M^4_{Pn}}{(m_1+m_2)^2} \delta(s-M_{Pn}^2)\;,\label{29}
\end{eqnarray}
The integrals in eqs.(\ref{28})-(\ref{29}) are simply calculated, and
this procedure is generally used.

In the presented scheme, let us introduce the function of state number
$n(s)$, so that
\begin{equation}
n(m_k^2) = k\;.
\end{equation}
This definition seems to be reasonable in the resonant region.
Then one has, for example, that
\begin{equation}
\frac{1}{\pi}\; \Im m \Pi_V^{(res)}(s) = s f^2_{Vn(s)}\; \frac{d}{ds} \sum_k
\theta(s-M^2_{Vk})\;.
\end{equation}
Further, it is evident that
\begin{equation}
\frac{d}{ds} \sum_k \theta(s-M_k^2) = \frac{dn(s)}{ds}\;\frac{d}{dn} \sum_k
\theta(n-k)\;,
\end{equation}
and eq.(\ref{28}) can be rewritten as
\begin{equation}
\Pi_V^{(res)}(q^2) = \int \frac{ds}{s-q^2}\; s f^2_{Vn(s)}\;\frac{dn(s)}{ds}\;
\frac{d}{dn} \sum_k \theta(n-k)\;.
\end{equation}
The "smooth average value" scheme means that
\begin{equation}
\Pi_V^{(res)}(q^2) = <\frac{d}{dn} \sum_k \theta(n-k)>\; \int \frac{ds}{s-q^2}
s f^2_{Vn(s)} \frac{dn(s)}{ds}\;.
\end{equation}
It is evident that, in average, the first derivative of step-like function
in the resonant region is equal to
\begin{equation}
<\frac{d}{dn} \sum_k \theta(n-k)> \simeq 1\;.
\end{equation}
Thus, in the scheme one has
\begin{eqnarray}
<\Pi_V^{(res)}(q^2)> & \approx & \int \frac{ds}{s-q^2}
s f^2_{Vn(s)}\; \frac{dn(s)}{ds}\;,
\label{35} \\
<\Pi_P^{(res)}(q^2)> & \approx & \int \frac{ds}{s-q^2}
\frac{s^2 f^2_{Pn(s)}}{(m_1+m_2)^2}\; \frac{dn(s)}{ds}\;.
\label{36}
\end{eqnarray}
Eqs.(\ref{35})-(\ref{36}) give the average correlators for the vector and
pseudoscalar mesons, therefore, due to eq.(\ref{21}) we state  that
\begin{equation}
\Im m <\Pi^{(hadr)}(q^2)> = \Im m \Pi^{(QCD)}(q^2)\;,
\end{equation}
that gives with account of eqs.(\ref{27}), (\ref{35}) and
(\ref{36}) at the physical points $s_n =M_n^2$
\begin{equation}
\frac{f_n^2}{M_n} = \frac{\alpha_S}{\pi} \; \frac{dM_n}{dn}
\; \biggl(\frac{4\mu}{M}\biggr)^2\;, \label{38}
\end{equation}
where in the limit of heavy quarks we use, that for the resonances
one has
\begin{equation}
m_1 +m_2 \approx M\;,\label{39}
\end{equation}
so that
\begin{equation}
f_{Vn} \simeq f_{Pn} = f_n\;.\label{40}
\end{equation}
Thus, one can conclude that for the heavy quarkonia the QCD sum rules give
the identity of $f_P$ and $f_V$ values for the pseudoscalar
and vector states.

Eq.(\ref{38}) differs from the ordinary sum rule scheme because it does not
contain the parameters, which are external to QCD. The quantity
$dM_n/dn$ is purely phenomenological. It defines the average mass difference
between the nearest levels with the identical quantum numbers.

Further, as it has been shown in ref.\cite{6}, in the region of average
distances between the heavy quarks in the charmonium and the bottomonium,
\begin{equation}
0.1\; fm < r < 1\;fm\;, \label{2.1}
\end{equation}
the QCD-motivated potentials allow the approximation in the form of
logarithmic law \cite{7} with the simple scaling properties, so
\begin{equation}
\frac{dn}{dM_n} = const.\;,\label{2.2}
\end{equation}
i.e. the density of heavy quarkonium states with the given quantum
numbers do not depend on the heavy quark flavours.

In ref.\cite{4} it has been shown, that relation (\ref{2.2}) is also
practically valid for the heavy quark potential approximation by the
power law (Martin potential) \cite{8}, where, neglecting a low value of
the binding energy for the quarks inside the quarkonium, one can again
get eq.(\ref{2.2}).

In ref.\cite{3} it has been found, that relation (\ref{2.2}) is valid
 with the accuracy up to small logarithmic corrections over the
reduced mass of quarkonium, if one makes the quantization of
$S$-wave states for the quarkonium with the Martin potential by the
Bohr-Sommerfeld procedure.

Moreover, with the accuracy up to the logarithmic corrections, $\alpha_S$
is the constant value. Thus, as it has been shown in refs.\cite{3,4},
for the leptonic constants of $S$-wave quarkonia, the scaling relation
takes place
\begin{equation}
\frac{f^2}{M}\; \biggl(\frac{M}{4\mu}\biggr)^2 = const.\;, \label{2.3}
\end{equation}
independently of the heavy quark flavours.

Taking into the account eqs.(\ref{39}) and (\ref{40}) and integrating
eqs.(\ref{35}), (\ref{36}) by parts, one can get with the accuracy up to
border terms, that one has
\begin{equation}
-2f_n\; \frac{df_n}{dn}\; \frac{dn}{dM_n}\; n = \frac{\alpha_s}{\pi}\;
M_n\; \biggl(\frac{4\mu}{M_n}\biggr)^2\;. \label{2.4}
\end{equation}
Comparing eqs.(\ref{38}) and (\ref{2.4}), one finds
\begin{equation}
\frac{df_n}{f_n dn} = - \frac{1}{2n}\;, \label{2.5}
\end{equation}
that  gives, after the integration,
\begin{equation}
\frac{f^2_{n_1}}{f^2_{n_2}} = \frac{n_2}{n_1}\;. \label{2.6}
\end{equation}
Relation (\ref{2.6}) leads to that the border terms, which have been neglected
in the writing of eq.(\ref{2.4}), are identically equal to zero.

First, note that eq.(\ref{2.3}), relating the leptonic constants of
different quarkonia, turns out to be certainly valid
for the quarkonia with the hidden
flavour ($c\bar c$, $b\bar b$), where $4\mu/M=1$ (see \cite{3,4}).

Second, eq.(\ref{2.3}) gives estimates of the leptonic constants for
the heavy $B$ and $D$ mesons, so these estimates are in a good agreement with
the values, obtained in the framework of other schemes of the QCD
sum rules \cite{4}.

Third, taking a value of the $1S$-level leptonic constant as the input one, we
have calculated the leptonic constants of higher $nS$-excitations in the
charmonium and the bottomonium and found a good agreement with the experimental
values \cite{5}.

These three facts show that the offered scheme can be quite reliably applied
to the systems with the heavy quarks.

Further, from eqs.(\ref{38}) and (\ref{2.6}) it follows that
\begin{equation}
\frac{f_1^2}{n}\; \frac{1}{M_n} = \frac{\alpha_S}{\pi}\;
\biggl(\frac{4\mu}{M_n}\biggr)^2\; \frac{dM_n}{dn}\;, \label{2.7}
\end{equation}
so that, neglecting the low value of quark binding energy
($M_n=M_1(1+O(1/M))$), one gets
\begin{equation}
\frac{dM_n}{dn}= \frac{1}{n}\; \frac{dM_n}{dn}(n=1)\;. \label{2.8}
\end{equation}
Integrating eq.(\ref{2.8}), one partially finds eq.(\ref{m3})
\begin{equation}
\frac{M_n-M_1}{M_2-M_1} = \frac{\ln {n}}{\ln {2}}\;,\;\;\;n\ge 2\;,\label{2.9}
\end{equation}
and
\begin{equation}
{M_2-M_1} = \frac{dM_n}{dn}(n=1)\; {\ln {2}}\;.\label{2.10}
\end{equation}

Thus, in the offered scheme of QCD sum rules, one takes into account the
Coulomb-like $\alpha_S/v$-corrections and, neglecing the power corrections over
the inverse heavy quark mass, one gets the universal relation for the
differences of $nS$-wave level masses of the heavy quarkonium.
\setlength{\unitlength}{0.85mm}\thicklines
\begin{figure}[t]
\begin{center}
\begin{picture}(100,90)
\put(15,10){\framebox(60,70)}
\put(3,10){$0$}
\put(15,30){\line(1,0){2}}
\put(3,30){$1$}
\put(15,50){\line(1,0){2}}
\put(3,50){$2$}
\put(15,70){\line(1,0){2}}
\put(3,70){$3$}

\put(15,20){\line(1,0){2}}
\put(15,40){\line(1,0){2}}
\put(15,60){\line(1,0){2}}

\put(0,83){$\alpha(n)$}

\put(25,10){\line(0,1){2}}
\put(35,10){\line(0,1){2}}
\put(45,10){\line(0,1){2}}
\put(55,10){\line(0,1){2}}
\put(65,10){\line(0,1){2}}
\put(15,2){$1$}
\put(25,2){$2$}
\put(35,2){$3$}
\put(45,2){$4$}
\put(55,2){$5$}
\put(65,2){$6$}
\put(75,2){$n$}

\put(15,10){\circle*{1.6}}
\put(25,30){\circle*{1.6}}
\put(35,41.8){\circle*{1.6}}
\put(45,49.8){\circle*{1.6}}
\put(55,60){\circle*{1.6}}
\put(65,65.6){\circle*{1.6}}

\put(14,09){\framebox(2,2)}
\put(24,29){\framebox(2,2)}
\put(34,41){\framebox(2,2)}
\put(44,45){\framebox(2,2)}
\put(54,53.8){\framebox(2,2)}
%put(35,22){\circle{1.6}}
%put(45,26){\circle{1.6}}
%put(55,34.8){\circle{1.6}}

%\input{mf.dat}
 \put( 15.10, 10.29){\circle*{0.75}}
 \put( 15.20, 10.57){\circle*{0.75}}
 \put( 15.30, 10.85){\circle*{0.75}}
 \put( 15.40, 11.13){\circle*{0.75}}
 \put( 15.50, 11.41){\circle*{0.75}}
 \put( 15.60, 11.68){\circle*{0.75}}
 \put( 15.70, 11.95){\circle*{0.75}}
 \put( 15.80, 12.22){\circle*{0.75}}
 \put( 15.90, 12.49){\circle*{0.75}}
 \put( 16.00, 12.75){\circle*{0.75}}
 \put( 16.10, 13.01){\circle*{0.75}}
 \put( 16.20, 13.27){\circle*{0.75}}
 \put( 16.30, 13.53){\circle*{0.75}}
 \put( 16.40, 13.78){\circle*{0.75}}
 \put( 16.50, 14.03){\circle*{0.75}}
 \put( 16.60, 14.28){\circle*{0.75}}
 \put( 16.70, 14.53){\circle*{0.75}}
 \put( 16.80, 14.78){\circle*{0.75}}
 \put( 16.90, 15.02){\circle*{0.75}}
 \put( 17.00, 15.26){\circle*{0.75}}
 \put( 17.10, 15.50){\circle*{0.75}}
 \put( 17.20, 15.74){\circle*{0.75}}
 \put( 17.30, 15.97){\circle*{0.75}}
 \put( 17.40, 16.21){\circle*{0.75}}
 \put( 17.50, 16.44){\circle*{0.75}}
 \put( 17.60, 16.67){\circle*{0.75}}
 \put( 17.70, 16.90){\circle*{0.75}}
 \put( 17.80, 17.12){\circle*{0.75}}
 \put( 17.90, 17.35){\circle*{0.75}}
 \put( 18.00, 17.57){\circle*{0.75}}
 \put( 18.10, 17.79){\circle*{0.75}}
 \put( 18.20, 18.01){\circle*{0.75}}
 \put( 18.30, 18.23){\circle*{0.75}}
 \put( 18.40, 18.44){\circle*{0.75}}
 \put( 18.50, 18.66){\circle*{0.75}}
 \put( 18.60, 18.87){\circle*{0.75}}
 \put( 18.70, 19.08){\circle*{0.75}}
 \put( 18.80, 19.29){\circle*{0.75}}
 \put( 18.90, 19.50){\circle*{0.75}}
 \put( 19.00, 19.71){\circle*{0.75}}
 \put( 19.10, 19.91){\circle*{0.75}}
 \put( 19.20, 20.12){\circle*{0.75}}
 \put( 19.30, 20.32){\circle*{0.75}}
 \put( 19.40, 20.52){\circle*{0.75}}
 \put( 19.50, 20.72){\circle*{0.75}}
 \put( 19.60, 20.92){\circle*{0.75}}
 \put( 19.70, 21.12){\circle*{0.75}}
 \put( 19.80, 21.31){\circle*{0.75}}
 \put( 19.90, 21.51){\circle*{0.75}}
 \put( 20.00, 21.70){\circle*{0.75}}
 \put( 20.10, 21.89){\circle*{0.75}}
 \put( 20.20, 22.08){\circle*{0.75}}
 \put( 20.30, 22.27){\circle*{0.75}}
 \put( 20.40, 22.46){\circle*{0.75}}
 \put( 20.50, 22.65){\circle*{0.75}}
 \put( 20.60, 22.83){\circle*{0.75}}
 \put( 20.70, 23.02){\circle*{0.75}}
 \put( 20.80, 23.20){\circle*{0.75}}
 \put( 20.90, 23.38){\circle*{0.75}}
 \put( 21.00, 23.56){\circle*{0.75}}
 \put( 21.10, 23.74){\circle*{0.75}}
 \put( 21.20, 23.92){\circle*{0.75}}
 \put( 21.30, 24.10){\circle*{0.75}}
 \put( 21.40, 24.27){\circle*{0.75}}
 \put( 21.50, 24.45){\circle*{0.75}}
 \put( 21.60, 24.62){\circle*{0.75}}
 \put( 21.70, 24.80){\circle*{0.75}}
 \put( 21.80, 24.97){\circle*{0.75}}
 \put( 21.90, 25.14){\circle*{0.75}}
 \put( 22.00, 25.31){\circle*{0.75}}
 \put( 22.10, 25.48){\circle*{0.75}}
 \put( 22.20, 25.65){\circle*{0.75}}
 \put( 22.30, 25.82){\circle*{0.75}}
 \put( 22.40, 25.98){\circle*{0.75}}
 \put( 22.50, 26.15){\circle*{0.75}}
 \put( 22.60, 26.31){\circle*{0.75}}
 \put( 22.70, 26.47){\circle*{0.75}}
 \put( 22.80, 26.64){\circle*{0.75}}
 \put( 22.90, 26.80){\circle*{0.75}}
 \put( 23.00, 26.96){\circle*{0.75}}
 \put( 23.10, 27.12){\circle*{0.75}}
 \put( 23.20, 27.28){\circle*{0.75}}
 \put( 23.30, 27.44){\circle*{0.75}}
 \put( 23.40, 27.59){\circle*{0.75}}
 \put( 23.50, 27.75){\circle*{0.75}}
 \put( 23.60, 27.91){\circle*{0.75}}
 \put( 23.70, 28.06){\circle*{0.75}}
 \put( 23.80, 28.21){\circle*{0.75}}
 \put( 23.90, 28.37){\circle*{0.75}}
 \put( 24.00, 28.52){\circle*{0.75}}
 \put( 24.10, 28.67){\circle*{0.75}}
 \put( 24.20, 28.82){\circle*{0.75}}
 \put( 24.30, 28.97){\circle*{0.75}}
 \put( 24.40, 29.12){\circle*{0.75}}
 \put( 24.50, 29.27){\circle*{0.75}}
 \put( 24.60, 29.42){\circle*{0.75}}
 \put( 24.70, 29.56){\circle*{0.75}}
 \put( 24.80, 29.71){\circle*{0.75}}
 \put( 24.90, 29.86){\circle*{0.75}}
 \put( 25.00, 30.00){\circle*{0.75}}
 \put( 25.10, 30.14){\circle*{0.75}}
 \put( 25.20, 30.29){\circle*{0.75}}
 \put( 25.30, 30.43){\circle*{0.75}}
 \put( 25.40, 30.57){\circle*{0.75}}
 \put( 25.50, 30.71){\circle*{0.75}}
 \put( 25.60, 30.85){\circle*{0.75}}
 \put( 25.70, 30.99){\circle*{0.75}}
 \put( 25.80, 31.13){\circle*{0.75}}
 \put( 25.90, 31.27){\circle*{0.75}}
 \put( 26.00, 31.41){\circle*{0.75}}
 \put( 26.10, 31.54){\circle*{0.75}}
 \put( 26.20, 31.68){\circle*{0.75}}
 \put( 26.30, 31.82){\circle*{0.75}}
 \put( 26.40, 31.95){\circle*{0.75}}
 \put( 26.50, 32.09){\circle*{0.75}}
 \put( 26.60, 32.22){\circle*{0.75}}
 \put( 26.70, 32.35){\circle*{0.75}}
 \put( 26.80, 32.49){\circle*{0.75}}
 \put( 26.90, 32.62){\circle*{0.75}}
 \put( 27.00, 32.75){\circle*{0.75}}
 \put( 27.10, 32.88){\circle*{0.75}}
 \put( 27.20, 33.01){\circle*{0.75}}
 \put( 27.30, 33.14){\circle*{0.75}}
 \put( 27.40, 33.27){\circle*{0.75}}
 \put( 27.50, 33.40){\circle*{0.75}}
 \put( 27.60, 33.53){\circle*{0.75}}
 \put( 27.70, 33.65){\circle*{0.75}}
 \put( 27.80, 33.78){\circle*{0.75}}
 \put( 27.90, 33.91){\circle*{0.75}}
 \put( 28.00, 34.03){\circle*{0.75}}
 \put( 28.10, 34.16){\circle*{0.75}}
 \put( 28.20, 34.28){\circle*{0.75}}
 \put( 28.30, 34.41){\circle*{0.75}}
 \put( 28.40, 34.53){\circle*{0.75}}
 \put( 28.50, 34.65){\circle*{0.75}}
 \put( 28.60, 34.78){\circle*{0.75}}
 \put( 28.70, 34.90){\circle*{0.75}}
 \put( 28.80, 35.02){\circle*{0.75}}
 \put( 28.90, 35.14){\circle*{0.75}}
 \put( 29.00, 35.26){\circle*{0.75}}
 \put( 29.10, 35.38){\circle*{0.75}}
 \put( 29.20, 35.50){\circle*{0.75}}
 \put( 29.30, 35.62){\circle*{0.75}}
 \put( 29.40, 35.74){\circle*{0.75}}
 \put( 29.50, 35.86){\circle*{0.75}}
 \put( 29.60, 35.97){\circle*{0.75}}
 \put( 29.70, 36.09){\circle*{0.75}}
 \put( 29.80, 36.21){\circle*{0.75}}
 \put( 29.90, 36.32){\circle*{0.75}}
 \put( 30.00, 36.44){\circle*{0.75}}
 \put( 30.10, 36.55){\circle*{0.75}}
 \put( 30.20, 36.67){\circle*{0.75}}
 \put( 30.30, 36.78){\circle*{0.75}}
 \put( 30.40, 36.90){\circle*{0.75}}
 \put( 30.50, 37.01){\circle*{0.75}}
 \put( 30.60, 37.12){\circle*{0.75}}
 \put( 30.70, 37.24){\circle*{0.75}}
 \put( 30.80, 37.35){\circle*{0.75}}
 \put( 30.90, 37.46){\circle*{0.75}}
 \put( 31.00, 37.57){\circle*{0.75}}
 \put( 31.10, 37.68){\circle*{0.75}}
 \put( 31.20, 37.79){\circle*{0.75}}
 \put( 31.30, 37.90){\circle*{0.75}}
 \put( 31.40, 38.01){\circle*{0.75}}
 \put( 31.50, 38.12){\circle*{0.75}}
 \put( 31.60, 38.23){\circle*{0.75}}
 \put( 31.70, 38.34){\circle*{0.75}}
 \put( 31.80, 38.44){\circle*{0.75}}
 \put( 31.90, 38.55){\circle*{0.75}}
 \put( 32.00, 38.66){\circle*{0.75}}
 \put( 32.10, 38.77){\circle*{0.75}}
 \put( 32.20, 38.87){\circle*{0.75}}
 \put( 32.30, 38.98){\circle*{0.75}}
 \put( 32.40, 39.08){\circle*{0.75}}
 \put( 32.50, 39.19){\circle*{0.75}}
 \put( 32.60, 39.29){\circle*{0.75}}
 \put( 32.70, 39.40){\circle*{0.75}}
 \put( 32.80, 39.50){\circle*{0.75}}
 \put( 32.90, 39.61){\circle*{0.75}}
 \put( 33.00, 39.71){\circle*{0.75}}
 \put( 33.10, 39.81){\circle*{0.75}}
 \put( 33.20, 39.91){\circle*{0.75}}
 \put( 33.30, 40.02){\circle*{0.75}}
 \put( 33.40, 40.12){\circle*{0.75}}
 \put( 33.50, 40.22){\circle*{0.75}}
 \put( 33.60, 40.32){\circle*{0.75}}
 \put( 33.70, 40.42){\circle*{0.75}}
 \put( 33.80, 40.52){\circle*{0.75}}
 \put( 33.90, 40.62){\circle*{0.75}}
 \put( 34.00, 40.72){\circle*{0.75}}
 \put( 34.10, 40.82){\circle*{0.75}}
 \put( 34.20, 40.92){\circle*{0.75}}
 \put( 34.30, 41.02){\circle*{0.75}}
 \put( 34.40, 41.12){\circle*{0.75}}
 \put( 34.50, 41.21){\circle*{0.75}}
 \put( 34.60, 41.31){\circle*{0.75}}
 \put( 34.70, 41.41){\circle*{0.75}}
 \put( 34.80, 41.51){\circle*{0.75}}
 \put( 34.90, 41.60){\circle*{0.75}}
 \put( 35.00, 41.70){\circle*{0.75}}
 \put( 35.10, 41.80){\circle*{0.75}}
 \put( 35.20, 41.89){\circle*{0.75}}
 \put( 35.30, 41.99){\circle*{0.75}}
 \put( 35.40, 42.08){\circle*{0.75}}
 \put( 35.50, 42.18){\circle*{0.75}}
 \put( 35.60, 42.27){\circle*{0.75}}
 \put( 35.70, 42.36){\circle*{0.75}}
 \put( 35.80, 42.46){\circle*{0.75}}
 \put( 35.90, 42.55){\circle*{0.75}}
 \put( 36.00, 42.65){\circle*{0.75}}
 \put( 36.10, 42.74){\circle*{0.75}}
 \put( 36.20, 42.83){\circle*{0.75}}
 \put( 36.30, 42.92){\circle*{0.75}}
 \put( 36.40, 43.02){\circle*{0.75}}
 \put( 36.50, 43.11){\circle*{0.75}}
 \put( 36.60, 43.20){\circle*{0.75}}
 \put( 36.70, 43.29){\circle*{0.75}}
 \put( 36.80, 43.38){\circle*{0.75}}
 \put( 36.90, 43.47){\circle*{0.75}}
 \put( 37.00, 43.56){\circle*{0.75}}
 \put( 37.10, 43.65){\circle*{0.75}}
 \put( 37.20, 43.74){\circle*{0.75}}
 \put( 37.30, 43.83){\circle*{0.75}}
 \put( 37.40, 43.92){\circle*{0.75}}
 \put( 37.50, 44.01){\circle*{0.75}}
 \put( 37.60, 44.10){\circle*{0.75}}
 \put( 37.70, 44.19){\circle*{0.75}}
 \put( 37.80, 44.27){\circle*{0.75}}
 \put( 37.90, 44.36){\circle*{0.75}}
 \put( 38.00, 44.45){\circle*{0.75}}
 \put( 38.10, 44.54){\circle*{0.75}}
 \put( 38.20, 44.62){\circle*{0.75}}
 \put( 38.30, 44.71){\circle*{0.75}}
 \put( 38.40, 44.80){\circle*{0.75}}
 \put( 38.50, 44.88){\circle*{0.75}}
 \put( 38.60, 44.97){\circle*{0.75}}
 \put( 38.70, 45.05){\circle*{0.75}}
 \put( 38.80, 45.14){\circle*{0.75}}
 \put( 38.90, 45.23){\circle*{0.75}}
 \put( 39.00, 45.31){\circle*{0.75}}
 \put( 39.10, 45.40){\circle*{0.75}}
 \put( 39.20, 45.48){\circle*{0.75}}
 \put( 39.30, 45.56){\circle*{0.75}}
 \put( 39.40, 45.65){\circle*{0.75}}
 \put( 39.50, 45.73){\circle*{0.75}}
 \put( 39.60, 45.82){\circle*{0.75}}
 \put( 39.70, 45.90){\circle*{0.75}}
 \put( 39.80, 45.98){\circle*{0.75}}
 \put( 39.90, 46.06){\circle*{0.75}}
 \put( 40.00, 46.15){\circle*{0.75}}
 \put( 40.10, 46.23){\circle*{0.75}}
 \put( 40.20, 46.31){\circle*{0.75}}
 \put( 40.30, 46.39){\circle*{0.75}}
 \put( 40.40, 46.47){\circle*{0.75}}
 \put( 40.50, 46.56){\circle*{0.75}}
 \put( 40.60, 46.64){\circle*{0.75}}
 \put( 40.70, 46.72){\circle*{0.75}}
 \put( 40.80, 46.80){\circle*{0.75}}
 \put( 40.90, 46.88){\circle*{0.75}}
 \put( 41.00, 46.96){\circle*{0.75}}
 \put( 41.10, 47.04){\circle*{0.75}}
 \put( 41.20, 47.12){\circle*{0.75}}
 \put( 41.30, 47.20){\circle*{0.75}}
 \put( 41.40, 47.28){\circle*{0.75}}
 \put( 41.50, 47.36){\circle*{0.75}}
 \put( 41.60, 47.44){\circle*{0.75}}
 \put( 41.70, 47.52){\circle*{0.75}}
 \put( 41.80, 47.59){\circle*{0.75}}
 \put( 41.90, 47.67){\circle*{0.75}}
 \put( 42.00, 47.75){\circle*{0.75}}
 \put( 42.10, 47.83){\circle*{0.75}}
 \put( 42.20, 47.91){\circle*{0.75}}
 \put( 42.30, 47.98){\circle*{0.75}}
 \put( 42.40, 48.06){\circle*{0.75}}
 \put( 42.50, 48.14){\circle*{0.75}}
 \put( 42.60, 48.21){\circle*{0.75}}
 \put( 42.70, 48.29){\circle*{0.75}}
 \put( 42.80, 48.37){\circle*{0.75}}
 \put( 42.90, 48.44){\circle*{0.75}}
 \put( 43.00, 48.52){\circle*{0.75}}
 \put( 43.10, 48.60){\circle*{0.75}}
 \put( 43.20, 48.67){\circle*{0.75}}
 \put( 43.30, 48.75){\circle*{0.75}}
 \put( 43.40, 48.82){\circle*{0.75}}
 \put( 43.50, 48.90){\circle*{0.75}}
 \put( 43.60, 48.97){\circle*{0.75}}
 \put( 43.70, 49.05){\circle*{0.75}}
 \put( 43.80, 49.12){\circle*{0.75}}
 \put( 43.90, 49.20){\circle*{0.75}}
 \put( 44.00, 49.27){\circle*{0.75}}
 \put( 44.10, 49.34){\circle*{0.75}}
 \put( 44.20, 49.42){\circle*{0.75}}
 \put( 44.30, 49.49){\circle*{0.75}}
 \put( 44.40, 49.56){\circle*{0.75}}
 \put( 44.50, 49.64){\circle*{0.75}}
 \put( 44.60, 49.71){\circle*{0.75}}
 \put( 44.70, 49.78){\circle*{0.75}}
 \put( 44.80, 49.86){\circle*{0.75}}
 \put( 44.90, 49.93){\circle*{0.75}}
 \put( 45.00, 50.00){\circle*{0.75}}
 \put( 45.10, 50.07){\circle*{0.75}}
 \put( 45.20, 50.14){\circle*{0.75}}
 \put( 45.30, 50.22){\circle*{0.75}}
 \put( 45.40, 50.29){\circle*{0.75}}
 \put( 45.50, 50.36){\circle*{0.75}}
 \put( 45.60, 50.43){\circle*{0.75}}
 \put( 45.70, 50.50){\circle*{0.75}}
 \put( 45.80, 50.57){\circle*{0.75}}
 \put( 45.90, 50.64){\circle*{0.75}}
 \put( 46.00, 50.71){\circle*{0.75}}
 \put( 46.10, 50.78){\circle*{0.75}}
 \put( 46.20, 50.85){\circle*{0.75}}
 \put( 46.30, 50.92){\circle*{0.75}}
 \put( 46.40, 50.99){\circle*{0.75}}
 \put( 46.50, 51.06){\circle*{0.75}}
 \put( 46.60, 51.13){\circle*{0.75}}
 \put( 46.70, 51.20){\circle*{0.75}}
 \put( 46.80, 51.27){\circle*{0.75}}
 \put( 46.90, 51.34){\circle*{0.75}}
 \put( 47.00, 51.41){\circle*{0.75}}
 \put( 47.10, 51.48){\circle*{0.75}}
 \put( 47.20, 51.54){\circle*{0.75}}
 \put( 47.30, 51.61){\circle*{0.75}}
 \put( 47.40, 51.68){\circle*{0.75}}
 \put( 47.50, 51.75){\circle*{0.75}}
 \put( 47.60, 51.82){\circle*{0.75}}
 \put( 47.70, 51.88){\circle*{0.75}}
 \put( 47.80, 51.95){\circle*{0.75}}
 \put( 47.90, 52.02){\circle*{0.75}}
 \put( 48.00, 52.09){\circle*{0.75}}
 \put( 48.10, 52.15){\circle*{0.75}}
 \put( 48.20, 52.22){\circle*{0.75}}
 \put( 48.30, 52.29){\circle*{0.75}}
 \put( 48.40, 52.35){\circle*{0.75}}
 \put( 48.50, 52.42){\circle*{0.75}}
 \put( 48.60, 52.49){\circle*{0.75}}
 \put( 48.70, 52.55){\circle*{0.75}}
 \put( 48.80, 52.62){\circle*{0.75}}
 \put( 48.90, 52.68){\circle*{0.75}}
 \put( 49.00, 52.75){\circle*{0.75}}
 \put( 49.10, 52.82){\circle*{0.75}}
 \put( 49.20, 52.88){\circle*{0.75}}
 \put( 49.30, 52.95){\circle*{0.75}}
 \put( 49.40, 53.01){\circle*{0.75}}
 \put( 49.50, 53.08){\circle*{0.75}}
 \put( 49.60, 53.14){\circle*{0.75}}
 \put( 49.70, 53.21){\circle*{0.75}}
 \put( 49.80, 53.27){\circle*{0.75}}
 \put( 49.90, 53.33){\circle*{0.75}}
 \put( 50.00, 53.40){\circle*{0.75}}
 \put( 50.10, 53.46){\circle*{0.75}}
 \put( 50.20, 53.53){\circle*{0.75}}
 \put( 50.30, 53.59){\circle*{0.75}}
 \put( 50.40, 53.65){\circle*{0.75}}
 \put( 50.50, 53.72){\circle*{0.75}}
 \put( 50.60, 53.78){\circle*{0.75}}
 \put( 50.70, 53.84){\circle*{0.75}}
 \put( 50.80, 53.91){\circle*{0.75}}
 \put( 50.90, 53.97){\circle*{0.75}}
 \put( 51.00, 54.03){\circle*{0.75}}
 \put( 51.10, 54.10){\circle*{0.75}}
 \put( 51.20, 54.16){\circle*{0.75}}
 \put( 51.30, 54.22){\circle*{0.75}}
 \put( 51.40, 54.28){\circle*{0.75}}
 \put( 51.50, 54.34){\circle*{0.75}}
 \put( 51.60, 54.41){\circle*{0.75}}
 \put( 51.70, 54.47){\circle*{0.75}}
 \put( 51.80, 54.53){\circle*{0.75}}
 \put( 51.90, 54.59){\circle*{0.75}}
 \put( 52.00, 54.65){\circle*{0.75}}
 \put( 52.10, 54.71){\circle*{0.75}}
 \put( 52.20, 54.78){\circle*{0.75}}
 \put( 52.30, 54.84){\circle*{0.75}}
 \put( 52.40, 54.90){\circle*{0.75}}
 \put( 52.50, 54.96){\circle*{0.75}}
 \put( 52.60, 55.02){\circle*{0.75}}
 \put( 52.70, 55.08){\circle*{0.75}}
 \put( 52.80, 55.14){\circle*{0.75}}
 \put( 52.90, 55.20){\circle*{0.75}}
 \put( 53.00, 55.26){\circle*{0.75}}
 \put( 53.10, 55.32){\circle*{0.75}}
 \put( 53.20, 55.38){\circle*{0.75}}
 \put( 53.30, 55.44){\circle*{0.75}}
 \put( 53.40, 55.50){\circle*{0.75}}
 \put( 53.50, 55.56){\circle*{0.75}}
 \put( 53.60, 55.62){\circle*{0.75}}
 \put( 53.70, 55.68){\circle*{0.75}}
 \put( 53.80, 55.74){\circle*{0.75}}
 \put( 53.90, 55.80){\circle*{0.75}}
 \put( 54.00, 55.86){\circle*{0.75}}
 \put( 54.10, 55.91){\circle*{0.75}}
 \put( 54.20, 55.97){\circle*{0.75}}
 \put( 54.30, 56.03){\circle*{0.75}}
 \put( 54.40, 56.09){\circle*{0.75}}
 \put( 54.50, 56.15){\circle*{0.75}}
 \put( 54.60, 56.21){\circle*{0.75}}
 \put( 54.70, 56.26){\circle*{0.75}}
 \put( 54.80, 56.32){\circle*{0.75}}
 \put( 54.90, 56.38){\circle*{0.75}}
 \put( 55.00, 56.44){\circle*{0.75}}
 \put( 55.10, 56.50){\circle*{0.75}}
 \put( 55.20, 56.55){\circle*{0.75}}
 \put( 55.30, 56.61){\circle*{0.75}}
 \put( 55.40, 56.67){\circle*{0.75}}
 \put( 55.50, 56.73){\circle*{0.75}}
 \put( 55.60, 56.78){\circle*{0.75}}
 \put( 55.70, 56.84){\circle*{0.75}}
 \put( 55.80, 56.90){\circle*{0.75}}
 \put( 55.90, 56.95){\circle*{0.75}}
 \put( 56.00, 57.01){\circle*{0.75}}
 \put( 56.10, 57.07){\circle*{0.75}}
 \put( 56.20, 57.12){\circle*{0.75}}
 \put( 56.30, 57.18){\circle*{0.75}}
 \put( 56.40, 57.24){\circle*{0.75}}
 \put( 56.50, 57.29){\circle*{0.75}}
 \put( 56.60, 57.35){\circle*{0.75}}
 \put( 56.70, 57.40){\circle*{0.75}}
 \put( 56.80, 57.46){\circle*{0.75}}
 \put( 56.90, 57.51){\circle*{0.75}}
 \put( 57.00, 57.57){\circle*{0.75}}
 \put( 57.10, 57.63){\circle*{0.75}}
 \put( 57.20, 57.68){\circle*{0.75}}
 \put( 57.30, 57.74){\circle*{0.75}}
 \put( 57.40, 57.79){\circle*{0.75}}
 \put( 57.50, 57.85){\circle*{0.75}}
 \put( 57.60, 57.90){\circle*{0.75}}
 \put( 57.70, 57.96){\circle*{0.75}}
 \put( 57.80, 58.01){\circle*{0.75}}
 \put( 57.90, 58.07){\circle*{0.75}}
 \put( 58.00, 58.12){\circle*{0.75}}
 \put( 58.10, 58.17){\circle*{0.75}}
 \put( 58.20, 58.23){\circle*{0.75}}
 \put( 58.30, 58.28){\circle*{0.75}}
 \put( 58.40, 58.34){\circle*{0.75}}
 \put( 58.50, 58.39){\circle*{0.75}}
 \put( 58.60, 58.44){\circle*{0.75}}
 \put( 58.70, 58.50){\circle*{0.75}}
 \put( 58.80, 58.55){\circle*{0.75}}
 \put( 58.90, 58.61){\circle*{0.75}}
 \put( 59.00, 58.66){\circle*{0.75}}
 \put( 59.10, 58.71){\circle*{0.75}}
 \put( 59.20, 58.77){\circle*{0.75}}
 \put( 59.30, 58.82){\circle*{0.75}}
 \put( 59.40, 58.87){\circle*{0.75}}
 \put( 59.50, 58.93){\circle*{0.75}}
 \put( 59.60, 58.98){\circle*{0.75}}
 \put( 59.70, 59.03){\circle*{0.75}}
 \put( 59.80, 59.08){\circle*{0.75}}
 \put( 59.90, 59.14){\circle*{0.75}}
 \put( 60.00, 59.19){\circle*{0.75}}
 \put( 60.10, 59.24){\circle*{0.75}}
 \put( 60.20, 59.29){\circle*{0.75}}
 \put( 60.30, 59.35){\circle*{0.75}}
 \put( 60.40, 59.40){\circle*{0.75}}
 \put( 60.50, 59.45){\circle*{0.75}}
 \put( 60.60, 59.50){\circle*{0.75}}
 \put( 60.70, 59.55){\circle*{0.75}}
 \put( 60.80, 59.61){\circle*{0.75}}
 \put( 60.90, 59.66){\circle*{0.75}}
 \put( 61.00, 59.71){\circle*{0.75}}
 \put( 61.10, 59.76){\circle*{0.75}}
 \put( 61.20, 59.81){\circle*{0.75}}
 \put( 61.30, 59.86){\circle*{0.75}}
 \put( 61.40, 59.91){\circle*{0.75}}
 \put( 61.50, 59.97){\circle*{0.75}}
 \put( 61.60, 60.02){\circle*{0.75}}
 \put( 61.70, 60.07){\circle*{0.75}}
 \put( 61.80, 60.12){\circle*{0.75}}
 \put( 61.90, 60.17){\circle*{0.75}}
 \put( 62.00, 60.22){\circle*{0.75}}
 \put( 62.10, 60.27){\circle*{0.75}}
 \put( 62.20, 60.32){\circle*{0.75}}
 \put( 62.30, 60.37){\circle*{0.75}}
 \put( 62.40, 60.42){\circle*{0.75}}
 \put( 62.50, 60.47){\circle*{0.75}}
 \put( 62.60, 60.52){\circle*{0.75}}
 \put( 62.70, 60.57){\circle*{0.75}}
 \put( 62.80, 60.62){\circle*{0.75}}
 \put( 62.90, 60.67){\circle*{0.75}}
 \put( 63.00, 60.72){\circle*{0.75}}
 \put( 63.10, 60.77){\circle*{0.75}}
 \put( 63.20, 60.82){\circle*{0.75}}
 \put( 63.30, 60.87){\circle*{0.75}}
 \put( 63.40, 60.92){\circle*{0.75}}
 \put( 63.50, 60.97){\circle*{0.75}}
 \put( 63.60, 61.02){\circle*{0.75}}
 \put( 63.70, 61.07){\circle*{0.75}}
 \put( 63.80, 61.12){\circle*{0.75}}
 \put( 63.90, 61.17){\circle*{0.75}}
 \put( 64.00, 61.21){\circle*{0.75}}
 \put( 64.10, 61.26){\circle*{0.75}}
 \put( 64.20, 61.31){\circle*{0.75}}
 \put( 64.30, 61.36){\circle*{0.75}}
 \put( 64.40, 61.41){\circle*{0.75}}
 \put( 64.50, 61.46){\circle*{0.75}}
 \put( 64.60, 61.51){\circle*{0.75}}
 \put( 64.70, 61.55){\circle*{0.75}}
 \put( 64.80, 61.60){\circle*{0.75}}
 \put( 64.90, 61.65){\circle*{0.75}}
 \put( 65.00, 61.70){\circle*{0.75}}
 \put( 65.10, 61.75){\circle*{0.75}}
 \put( 65.20, 61.80){\circle*{0.75}}
 \put( 65.30, 61.84){\circle*{0.75}}
 \put( 65.40, 61.89){\circle*{0.75}}
 \put( 65.50, 61.94){\circle*{0.75}}
 \put( 65.60, 61.99){\circle*{0.75}}
 \put( 65.70, 62.03){\circle*{0.75}}
 \put( 65.80, 62.08){\circle*{0.75}}
 \put( 65.90, 62.13){\circle*{0.75}}
 \put( 66.00, 62.18){\circle*{0.75}}
 \put( 66.10, 62.22){\circle*{0.75}}
 \put( 66.20, 62.27){\circle*{0.75}}
 \put( 66.30, 62.32){\circle*{0.75}}
 \put( 66.40, 62.36){\circle*{0.75}}
 \put( 66.50, 62.41){\circle*{0.75}}
 \put( 66.60, 62.46){\circle*{0.75}}
 \put( 66.70, 62.51){\circle*{0.75}}
 \put( 66.80, 62.55){\circle*{0.75}}
 \put( 66.90, 62.60){\circle*{0.75}}
 \put( 67.00, 62.65){\circle*{0.75}}
 \put( 67.10, 62.69){\circle*{0.75}}
 \put( 67.20, 62.74){\circle*{0.75}}
 \put( 67.30, 62.78){\circle*{0.75}}
 \put( 67.40, 62.83){\circle*{0.75}}
 \put( 67.50, 62.88){\circle*{0.75}}
 \put( 67.60, 62.92){\circle*{0.75}}
 \put( 67.70, 62.97){\circle*{0.75}}
 \put( 67.80, 63.02){\circle*{0.75}}
 \put( 67.90, 63.06){\circle*{0.75}}
 \put( 68.00, 63.11){\circle*{0.75}}
 \put( 68.10, 63.15){\circle*{0.75}}
 \put( 68.20, 63.20){\circle*{0.75}}
 \put( 68.30, 63.24){\circle*{0.75}}
 \put( 68.40, 63.29){\circle*{0.75}}
 \put( 68.50, 63.34){\circle*{0.75}}
 \put( 68.60, 63.38){\circle*{0.75}}
 \put( 68.70, 63.43){\circle*{0.75}}
 \put( 68.80, 63.47){\circle*{0.75}}
 \put( 68.90, 63.52){\circle*{0.75}}
 \put( 69.00, 63.56){\circle*{0.75}}
 \put( 69.10, 63.61){\circle*{0.75}}
 \put( 69.20, 63.65){\circle*{0.75}}
 \put( 69.30, 63.70){\circle*{0.75}}
 \put( 69.40, 63.74){\circle*{0.75}}
 \put( 69.50, 63.79){\circle*{0.75}}
 \put( 69.60, 63.83){\circle*{0.75}}
 \put( 69.70, 63.88){\circle*{0.75}}
 \put( 69.80, 63.92){\circle*{0.75}}
 \put( 69.90, 63.96){\circle*{0.75}}
 \put( 70.00, 64.01){\circle*{0.75}}

\end{picture}
\end{center}
\caption{The experimental values of $nS$ - bottomonium (solid dots)
and charmonium (empty boxes) mass differences
$\alpha(n) =(M_n-M_1)/(M_2-M_1)$
and the dependence in the present model $\alpha(n)=\ln{n}/\ln{2}$.}
\label{fm}
\end{figure}
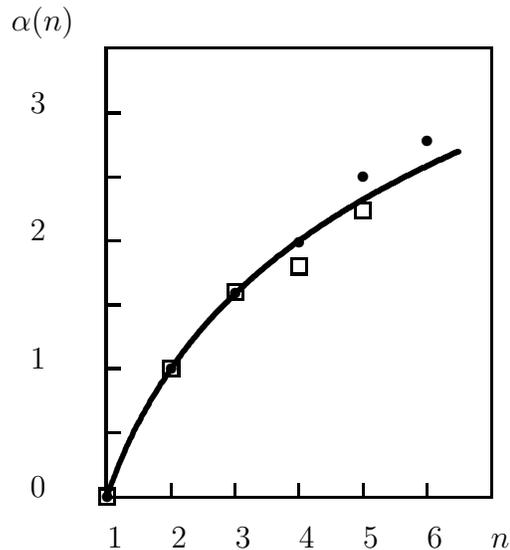

\section{Analysis of the Mass Difference Relation}

Eq.(\ref{2.9}) for the differences of $nS$-wave level masses of the heavy
quarkonium does not contain external parameters and it allows direct
comparison with the experimental data on the masses of particles in
the $\psi$- and $\Upsilon$-families \cite{PDG}.

Dependence (\ref{2.9}) and the experimental values for the relations of
heavy quarkonium masses are presented on Figure 1, where one neglects the
spin-spin splittings.

Note, the $\psi(3770)$ and $\psi(4040)$ charmonium states suppose to be the
results of the $3D$- and $3S$-states mixing, so that the $D$-wave dominates
in the $\psi(3770)$-state, and the mixing of the $3D$ and $3S$ wave functions
is accompanied by a small shifts of the masses, so that we have
supposed $M_3= M_{\psi(4040)}$.

As one can see from the Figure, relation (\ref{2.9}), obtained in the
leading approximation, is in a good agreement
with the experimental data\footnote{Note, that at $n\ge 4$, the $nS$-level
is above the threshold of decay into the heavy meson pair and the dynamics
of light quarks becomes essential. This dynamics is not taken into account
in the present model.}.

\section*{Conclusion}

In the framework of the specific scheme of QCD sum rules for the
two-point correlators of heavy quark currents, in the leading approximation
one takes into account the contribution of higher $nS$-levels in the heavy
quarkonium and derives the universal relation for the quarkonium mass
differences
$$
\frac{M_n-M_1}{M_2-M_1} = \frac{\ln {n}}{\ln {2}}\;,\;\;\;n\ge 2\;.
$$
This reflects the phenomenological flavour independence
of the kinetic energy in the bound states of heavy quarkonium.

\end{document}